# Evaluating Certificate Policy - Certification Practice Statement of Unique Government Certification Authority using Public Key Infrastructure Assessment Guidelines: Research in Progress

Dea Saka Kurnia Putra
Manajemen Persandian
Sekolah Tinggi Sandi Negara
Bogor, Indonesia
dea.saka@student.stsn-nci.ac.id

Edit Prima
Sekolah Tinggi Sandi Negara
Bogor, Indonesia
edit.prima@stsn-nci.ac.id

*Abstract*—**OSD PSE is the Indonesian Government's Certification Authority (CA) for National e-Procurement System and later named OSD PSE G2. It has a unique hierarchical structure under the OSD Lemsaneg. As an Issuing CA, the OSD PSE G2 publishes and guarantee the quality of Certificate Policy and Certification Practice Statement (CP-CPS) in order to gain the PKI user's trustworthy. In this article, we analyze the CP-CPS version 1.0 that published by OSD PSE G2. For this purpose, we apply the methodology of PKI Assessment Guidelines (PAG). The quality assessment of this CP-CPS, including its compliance to the related reference/standard, namely: CP OSD Lemsaneg v.1.1; RFC 3647; and CA Business Practice Disclosure Principle on Trust Service Principles and Criteria for Certification Authorities (BPDP-TSPCCA) version 2.0. We finally found that the CP-CPS version 1.0 does not comply with related standard and reference. Hence, the CP-CPS need to be updated following the current condition of OSD PSE G2.**

## I. Introduction

The development of electronic transactions stimulates demands for the use of public key cryptography system to support the authentication service and non-repudiation in every electronic transaction activity. Public key cryptography system is an aspect of the Public Key Infrastructure (PKI) [1] [2]. Nowadays, the component technologies of PKI such the use of public key cryptography and underlying systems to enable digital signatures, strong authentication, data integrity, non-repudiation, and confidentiality is most often discussed [3]. Application of PKI in Indonesia is used in e-procurement, e-banking and e-shopping along with the enactment of Information and Electronic Transactions Act (UU ITE) and Government Regulation on the Implementation of Electronic Transaction System (PP PSTE) [4].

There are 5 main components in the implementation of the PKI, the Certification Authority (CA), Registration Authority (RA), PKI Client, Digital Certificate and Certificate Distribution System or Repository [5]. CA is the most important component for a digital certificate issuer. Sustainability of the CA will be run in accordance with its purpose if there are at least four types of main documents, namely the Relying Party Agreements, Subscriber Agreements, Certification Practice Statement (CPS), and Certificate Policy (CP) [6].

Otoritas Sertifikat Digital Lembaga Sandi Negara (OSD Lemsaneg) is a CA which issues, distributes, and manages digital certificates with the government agency. OSD Lemsaneg have several certification services refer to the type of use and the guarantee level categories, among others: OSD PSE and OSD Layanan Universal (LU). OSD Lemsaneg is managed by the government agency called Balai Sertifikasi Elektronik (BSrE) that provide information security services for electronic documents [7].

OSD PSE has a key pair cryptoperiod of 5 years. According to the OSD PSE key pair's cryptoperiod, BSrE as the OSD PSE's government agency have to extend the OSD PSE's key pair cryptoperiod. Key ceremony is held on 2016 to generate OSD PSE G2, OSD LU K1, OSD LU K2, OSD LU K3, and OSD LU K4's key pair with 10 years of cryptoperiod [8]. Start from 2016 OSD PSE named OSD PSE G2.

OSD PSE G2 has a unique hierarchical structure as a subordinate CA under the OSD Lemsaneg. According to a common standard of PKI operation, OSD PSE G2 also publish CP-CPS in 2012. The CP-CPS was addressed to regulate the OSD PSE's certification practices. CP-CPS OSD PSE version 1.0. seems do not relevant anymore because the OSD PSE is now become OSD PSE G2. Since the CP-CPS were a most important PKI component, we conduct a research to assess the CP-CPS in order to measure its compliance to the PKI standard. In the next section, we will review the researches related to the CP-CPS in OSD environment.

## II. RELATED WORKS

We have analyzed 4 related researches. The first research is about how the draft of CP OSD Lemsaneg is constructed [9]. The second research is about the CA environmental controls assessment in OSD PSE using TSPCCA version 2.0 [10]. The third research is about the draft of CP OSD Lemsaneg's assessment according to each RFC 3647 provisions [11]. The last research which done in 2017 is about the draft of CPS OSD Layanan Universal Kelas 2's (OSD LU K2) assessment using PKI Assessment Guidelines version 1.0 (PAG version 1.0). OSD LU K2 is an issuing CA which hierarchically under the OSD Lemsaneg.

The first research goal is to propose CP OSD Lemsaneg's draft using Soft System Methodology (SSM) [9]. The second research goal is to evaluate the OSD PSE's operational compliance with CA environmental controls principle of TSPCCA version 2.0 and to know the impact of the current condition [10]. The third research goal is to assess the draft of CP OSD Lemsaneg compliance according to each provision in RFC 3647 using gap analysis technique [11]. The last research goal is to assess the CPS OSD LU K2's draft according to CP OSD Lemsaneg version 1.0 and BPDP-TSPCCA version 2.0 using PAG version 1.0 [12].

TSPCCA version 2.0 provides a framework for PKI assessor to assess the compliance of CA according to its standard and policy which regulate it hierarchically. This document replaces the previous version of namely AICPA/CICA WebTrust Program for Certification Authorities (WPCA) that was issued in August 2000. Unlike WPCA which was intended to be used by licensed WebTrust practitioners only, this TSPCCA version 2.0 is regarded as "open-source" and can be used to assess a PKI employment by any third-party service provider. CA also can use TSPCCA version 2.0 as a guidance to done self assessment in their PKI employment [6].

PAG version 1.0 provides a guideline for assessing PKI operations related to the technical, legal, business, and policy issues. This guideline can be used by any party which aims to assess a PKI deployment [13]. The PAG methodology that used in this research will be discussed in the next section.

## III. PKI ASSESSMENT GUIDELINES

The scope of this research is focused on evaluating the CP-CPS OSD PSE version 1.0. This research is conducted using PKI Assessment Guidelines version 1.0 (PAG version 1.0). The research consists of 3 instead of 5 phases, namely: Planning, Policy Assessment, CPS Review, Operational Effectiveness Verification, and Reporting. The research only done Planning, Policy Assessment, and CPS Review Phase.

During the Planning phase, the assessor shall understands all of the literature used in the assessment including the PKI policy and its operational. Other objectives of this phase are to: 1) establish communication with parties involved in the assessment; and 2) refine the scope of the assessment together with the PKI staff [13].

Next, the assessor assesses whether the target Certificate Policy (CP) is suitable for its intended purposes during Policy Assessment phase. CP is assessed using the international recommendation which is RFC 3647. The assessment can be done using audit checklist, gap analysis, observation, and interview. The result of Policy Assessment phase is categorized into categories based on their reason of uncompliance. Last, the assessment result is tabulated to help the PKI staff understands the assessment result in simplified form [13].

Then, in the CPS Review phase, the compliance of CPS is assessed towards its CP and standard related to the CPS [13] [14]. First, CPS's compliance is assessed towards the policy which is the CP. Next, CPS's compliance is assessed towards the standard which in here TSPCCA version 2.0 is the standard used to assess the CPS compliance based on CA Business Practices Disclosure Principle [15]. The result of the compliance assessment is categorized into several categories based on their reason and their condition. Last, the result of CPS Review phase is tabulated in simplified form to help the PKI staff understands the compliance assessment result.

## IV. HIERARCHICAL STRUCTURE OF OSD PSE G2

Otoritas Sertifikat Digital Pengadaan Barang/Jasa Secara Elektronik also known as OSD PSE is an Indonesia's government Certification Authority (CA) [16]. OSD PSE has specific purposes and unique infrastructure of issuing the certificates [7]. Since 16[th] of August 2016, OSD PSE named as OSD PSE G2. The differences between OSD PSE and OSD PSE G2 is shown in Table I.

TABLE I. THE DIFFERENCES BETWEEN OSD PSE AND OSD PSE G2

| No | Differentiator | OSD PSE | OSD PSE G2 |
|---|---|---|---|
| 1 | CA's private key storing module | Not Applicable | Hardware Security Module (HSM) |
| 2 | CA's key pair cryptoperiod | 5 years | 10 years |
| 3 | Key pair generator's algorithm | RSA 4096-bit | RSA 2048-bit |

The differences between OSD PSE and OSD PSE G2 relate to the security issues. The implementation of HSM in OSD PSE G2 shows that OSD PSE G2 notice the vulnerability lies in the CA's private key. The CA's key pair cryptoperiod is setup to 10 years according to the recommendation given by NIST [17].

The uniqueness of OSD PSE G2 rely on the issuing hierarchy and the main function of the CA. The Indonesia government CA hierarchy is unique. OSD Lemsaneg as the temporary root CA of other CAs only has the CP but not the CPS. Another uniqueness of this hierarchy, the 5 issuing CA which runs under OSD Lemsaneg does not have any CP, but they have the CPS. Ideally, both root and issuing CA must have both of them [18]. The issuing hierarchy of OSD PSE G2 is shown in Fig. 1.

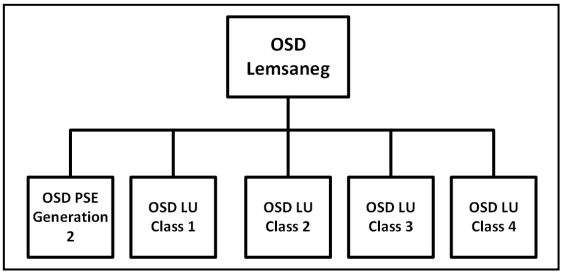

Fig. 1. Hierarchy of Indonesia Government Certification Authority

From the figure above, it is shown that OSD PSE G2 is an issuing CA along with OSD LU K1, OSD LU K2, OSD LU K3, and OSD LU K4. OSD PSE G2 is run under OSD Lemsaneg hierarchically which is the temporary root CA of the issuing CAs. Furthermore, the main function of OSD PSE G2 is in the scope of signs, issues, and maintains the certificates used in Sistem Pengadaan Secara Secara Elektronik also known as SPSE [16]. SPSE is a system which provides e-procurement participants a certification system in purpose to support the non-repudiation service. So, OSD PSE G2 only provides certification services in the scope of an e-procurement system which maintained Layanan Pengadaan Barang/Jasa Secara Elektronik (LPSE) also known as electronic procurement under Lembaga Kebijakan Pengadaan Barang/Jasa Pemerintah (LKPP) also known as institution of government procurement policy in Indonesia. LPSE is a service which provides a government electronic procurement system in Indonesia. In the next section, we will deeply discuss the analysis of CP-CPS published by OSD PSE G2.

## V. RESULT AND DISCUSSIONS

### A. Planning Phase

The goals of this phase are to establish communication with the parties involved in the CP-CPS PSE version 1.0 evaluation and to refine the scope of evaluation. Before establishing communication with the staff of OSD PSE G2, it needs to be clear that the documents related to the assessment are understood. The main literature needed to be understood for the assessment are: CP-CPS OSD PSE version 1.0, CP OSD Lemsaneg version 1.0, CP OSD Lemsaneg version 1.1, PKI Assessment Guidelines version 1.0, WebTrust Program for Certification Authorities version 1.0, Trust Service Principles and Criteria for Certification Authorities version 2.0, RFC 2527, and RFC 3647.

Hereinafter, establish communication with the OSD PSE G2 staff to explain the assessment's purpose and propose the assessment's scope. Focus Group Discussion (FGD) with the policy organization's staff is done to refine and adjust the assessment's scope proposed before. Finally, the assessment's scope is about to assess the CP-CPS OSD PSE version 1.0 using PAG version 1.0 based on CP OSD Lemsaneg version 1.1, RFC 3647, and CA BPDP-TSPCCA version 2.0.

### B. Policy Assessment Phase

In this phase, we review the compliance of CP OSD Lemsaneg v.1.1 towards RFC 3647 as recommended by [11] [12]. We conduct the gap analysis to find the gap between the recommendation and the CP OSD Lemsaneg version 1.1's current state. Further, the results of gap analysis can be used to refine the CP OSD Lemsaneg version 1.1. The gap analysis result is resumed and categorized into 7 categories show in Table II. Practically, the categorization is created after the gap analysis is done. The tabulation and categorization of the gap analysis result are shown in Table III.

TABLE II. CATEGORIZATION OF CP OSD LEMSANEG version 1.1 ASSESSMENT RESULT

| No | Category | Explanation |
|---|---|---|
| 1 | Category 1 | Recommendation will be implemented in the next revision of CP OSD Lemsaneg |
| 2 | Category 2 | Recommendation is unimplemented in CP OSD Lemsaneg version 1.1 |
| 3 | Category 3 | Outline of CP OSD Lemsaneg version 1.1's subchapters does not comply the provisions in RFC 3647. |
| 4 | Category 4 | Recommendation is stated "already implemented" by [12], but the fact it has not |
| 5 | Category 5 | CP OSD Lemsaneg version 1.1's outline is wrongly translated into Indonesian |
| 6 | Category 6 | Recommendation is not stated and analyzed in [12] |
| 7 | Category 7 | Recommendation is already implemented in CP OSD Lemsaneg version 1.1 |

From the result above, known that only 1 recommendation which is implemented in CP OSD Lemsaneg version 1.1 according to the evaluation done before [12].

TABLE III. TABULATION OF CP OSD LEMSANEG version 1.1 ASSESSMENT RESULT

| No | Category | Number of Criteria |
|---|---|---|
| 1 | Category 1 | 6 |
| 2 | Category 2 | 33 |
| 3 | Category 3 | 48 |
| 4 | Category 4 | 1 |
| 5 | Category 5 | 9 |
| 6 | Category 6 | 2 |
| 7 | Category 7 | 1 |

Based on the Table III, known that only 1 of 100 recommendation that already implemented in CP OSD Lemsaneg version 1.1. Category 1, 2, 3, 4, 5, and 6 are generally categorized as unfulfilled recommendation. So that, refinements on CP OSD Lemsaneg version 1.1 is needed.

### C. CPS Review Phase

CPS review phase is a phase where CP-CPS OSD PSE version 1.0 is assessed towards 3 documents and divided into 2 sub phases. The first sub phase is where the CP-CPS OSD PSE version 1.0 is assessed towards CP OSD Lemsaneg version 1.1 along with RFC 3647. The second sub phase is where the CP-CPS OSD PSE version 1.0 is assessed towards 45 criteria on CA BPDP-TSPCCA version 2.0. Below are the CP-CPS OSD PSE version 1.0 assessment towards those documents.

*1) Assessment Towards CP OSD Lemsaneg version 1.1*

During the assessment of CP-CPS OSD PSE version 1.0 towards CP OSD Lemsaneg version 1.1, the gap analysis results are categorized into 4 categories, namely Acceptable (A) for the statement in CP-CPS OSD PSE version 1.0 that comply the statement in CP OSD Lemsaneg version 1.1.

Statement categorized as Not Comparable (NC) when a statement in CP-CPS OSD PSE version 1.0 contains sub chapter contained in CP OSD Lemsaneg version 1.1 but the policy contains is dissimilar that makes it not comparable. Missing (M) when the CP-CPS OSD PSE version 1.0 does not contain detailed explanation about policy content or does not contain policy content at all according to CP OSD Lemsaneg version 1.1. Last, statement categorized as Not Applicable (NA) when the requirement contained within the CP OSD Lemsaneg version 1.1 is not transferable or applied with the CP-CPS OSD PSE version 1.1.

The method to assess CP-CPS OSD PSE version 1.0 towards CP OSD Lemsaneg version 1.1 is gap analysis. The gap analysis is done line-by-line between them. From 291 subchapters, it found that only 102 or 35.05% subchapters in CP-CPS OSD PSE version 1.0 comply to CP OSD Lemsaneg version 1.1 and categorized as Applicable (A). Hereinafter, it found that 21 or 7.22% subchapters in CP-CPS OSD PSE version 1.0 cannot be compared with CP OSD Lemsaneg version 1.1 and categorized as Not Comparable (NC). Next, it found that 91 or 31.27% subchapters in CP-CPS OSD PSE version 1.0 does not comply (or contain less explanation than) the statements in CP OSD Lemsaneg version 1.1. At last, 77 or 26.46% subchapters in CP OSD Lemsaneg version 1.1 cannot be applied to CP-CPS OSD PSE version 1.0 because of the statements stated in CP OSD Lemsaneg version 1.1 themselves do not comply the RFC 3647 provisions.

In this phase, the assessment result is categorized into more explained category based on the evidence found. The categorization table is shown in Table IV.

TABLE IV. CATEGORIZATION BY REASON OF CP-CPS OSD PSE VERSION 1.0 ASSESSMENT RESULT TOWARDS CP OSD LEMSANEG VERSION 1.1

| No | Category | Explanation |
|---|---|---|
| 1 | [ ] | Sub chapter contains no inconsistencies |
| 2 | [1] | Substantive mismatch |
| 3 | [2] | Typographic errors |
| 4 | [3] | Sub chapter is missing |
| 5 | [4] | Title of the sub chapter mismatch with RFC 3647 typographically |
| 6 | [5] | Title of the sub chapter mismatch with RFC 3647 which causes ambiguity |

The [ ] means that the sub chapter has no reason to be stated because it is categorized as Acceptable (A) which means CP-CPS OSD PSE version 1.0 is comply to the CP OSD Lemsaneg version 1.1 or Not Comparable (NC) which means CP-CPS OSD PSE version 1.0 contains subchapter which is not contained in CP OSD Lemsaneg version 1.1. A sub chapter can be categorized into one or more category based on its reason which makes it does not comply to the CP OSD Lemsaneg version 1.1. After knowing the categorization, the tabulation of assessment result can be found on Table V.

TABLE V. THE TABULATION OF CP-CPS OSD PSE VERSION 1.0 ASSESSMENT RESULT TOWARDS CP OSD LEMSANEG VERSION 1.1

| No | Category | Number of subchapter | | No | Category | Number of subchapter | |
|---|---|---|---|---|---|---|---|
| 1 | A | 22 | | 18 | M [1] | 18 | |
| 2 | A [2] | 9 | | 19 | M [3] | 46 | |
| 3 | A [3] | 1 | | 20 | M [1,2] | 3 | |
| 4 | A [4] | 41 | | 21 | M [1,3] | 1 | |
| 5 | A [5] | 9 | | 22 | M [1,4] | 9 | |
| 6 | A [1,4] | 1 | 102 | 23 | M [2,3] | 8 | 91 |
| 7 | A [2,4] | 16 | | 24 | M [1,2,4] | 4 | |
| 8 | A [2,5] | 1 | | 25 | M [1,2,5] | 1 | |
| 9 | A [1,2,4] | 1 | | 26 | M [2,3,4] | 1 | |
| 10 | NC | 7 | | 27 | NA [3] | 70 | |
| 11 | NC [2] | 1 | | 28 | NA [4] | 2 | |
| 12 | NC [4] | 4 | | 29 | NA [5] | 2 | 77 |
| 13 | NC [5] | 2 | | 30 | NA [1,2] | 1 | |
| 14 | NC [1,2] | 3 | 21 | 31 | NA [2,3] | 2 | |
| 15 | NC [1,4] | 2 | | | | | |
| 16 | NC [2,3] | 1 | | | | | |
| 17 | NC [2,4] | 1 | | | | | |
| | **Total: 291** | | | | | | |

Sub chapter categorized as Acceptable (A) is divided into 9 categories which the number of sub chapter categorized as (A) are 102. Not Comparable (NC) is divided into 8 categories and 21 sub chapters are categorized as (NC) totally. Next, Missing (M) is divided into 9 categories and 91 sub chapters are categrized as (M). Last, Not Applicable (NA) is divided into 5 categories and 77 sub chapters are categorized as (NA). From the assessment above, we can conclude that CP-CPS OSD PSE version 1.0 must be refined according to CP OSD Lemsaneg and RFC 3647's provisions.

*2) Assessment Towards CA BPDP-TSPCCA version 2.0*

In this sub phase, a gap analysis of CP-CPS OSD PSE version 1.0 is done towards criteria based on CA Business Practices Disclosure Principle on Trust Service Principles and Criteria for Certification Authorities version 2.0 (CA BPDP-TSPCCA version 2.0.) CA BPDP-TSPCCA version 2.0 is consisting of 45 criteria and 4 subprinciples namely: general; key life cycle management; certificate life cycle management; CA environmental controls. It found that 16 criteria are fulfilled by CP-CPS OSD PSE version 1.0 according to the BPDP-TSPCCA version 2.0.

Based on the gap analysis done, it found that CP-CPS OSD PSE version 1.0 complies only 16 of 45 criteria stated in CA BPDP-TSPCCA version 2.0. From 9 unfulfilled criteria in general, 4 criteria are fully unfulfilled and 5 criteria are partially unfulfilled. In key life cycle management found that 2 criteria are fully unfulfilled as well as partially unfulfilled. Forth, in certificate life cycle management can be found 5 criteria are fully unfulfilled and 5 criteria are partially unfulfilled. Lattermost, found that 2 criteria are fully unfulfilled and 4 criteria are partially unfulfilled in CA environmental controls subprinciple. The categorization of unfulfilled criteria is shown in Table VI.

TABLE VI. THE TABULATION OF CP-CPS OSD PSE VERSION 1.0 ASSESSMENT RESULT TOWARDS CA BPDP-TSPCCA VERSION 2.0

| No | Subprinciples | Number of Criteria | Fulfilled Criteria |
|---|---|---|---|
| 1 | General | 16 | 7 |
| 2 | Key Life Cycle Management | 7 | 3 |
| 3 | Certificate Life Cycle Management | 15 | 5 |
| 4 | CA Environmental Controls | 7 | 1 |
|  | Total | 45 | 16 |

From the categorization above, the known fact shows that the fully unfulfilled criteria is 13 and the partially unfulfilled criteria is 16 criteria in total. The total unfulfilled criteria is 29 of the total criteria. So, refinements on CP-CPS OSD PSE version 1.0 is needed to comply criteria on CA BPDP-TSPCCA version 2.0. So that, OSD PSE G2 can work well through the refinement. The categorization of unfulfilled criteria of CA BPDP-TSPCCA v2.0. is show on Table VII.

TABLE VII. THE CATEGORIZATION OF UNFULFILLED CRITERIAS IN CP-CPS OSD PSE VERSION 1.0 ASSESSMENT RESULT TOWARDS CA BPDP-TSPCCA V. 2.0

| No | Subprinciples | Fully Unfulfilled | Partially Unfulfilled |
|---|---|---|---|
| 1 | General | 5, 13, 15, 16 | 1, 6, 7, 9, 11 |
| 2 | Key Life Cycle Management | 19, 20 | 18, 23 |
| 3 | Certificate Life Cycle Management | 24, 28, 29, 34, 38 | 25, 26, 30, 33, 35 |
| 4 | CA Environmental Controls | 40, 42 | 39, 41, 43, 44 |
|  | Total | 13 | 16 |

From the table above, we know that some criteria have sub criteria which must be complied. It found that 13 unfulfilled criteria are fully unfulfilled and 16 criteria are partially unfulfilled by the CP-CPS OSD PSE version 1.0.

VI. CONCLUSION AND FUTURE WORKS

From the assessment using the gap analysis method on CP-CPS OSD PSE version 1.0 towards CP OSD Lemsaneg version 1.1, RFC 3647, and CA BPDP-TSPCCA version 2.0, we obtained that the CP-CPS OSD PSE version 1.0 is no longer relevant. It is proven by the assessment results that show only 102 out of 291 sub chapters in CP-CPS OSD PSE version 1.0 that comply the statements in CP OSD Lemsaneg version 1.1. Along with that, only 16 out of 45 criteria of CA Business Practices Disclosure Principle is complied by CP-CPS OSD PSE version 1.0. Hence, CP-CPS OSD PSE version 1.0 must be majority revised and name the document as CPS OSD PSE G2. Recommendation for the next research regarding is to continue the operational effectiveness verification and reporting phase. In the future, we will conduct Operational Effectiveness Verification and Reporting Phase to done evaluating the OSD PSE G2 operational and documentation.

ACKNOWLEDGMENT

This research was financially and non-financially supported by National Crypto Institute also known as Sekolah Tinggi Sandi Negara (STSN-NCI). We thank our colleagues who provided information, expertise, and insight and expertise that superbly assisted the research.